\documentclass[aps,reprint,nofootinbib,twocolumn,superscriptaddress]{revtex4-1}
\usepackage{subfigure}
\usepackage{amsmath, amssymb,graphicx,color,bm,epstopdf}
\usepackage[dvipsnames]{xcolor}
\usepackage{bbold}
\usepackage{braket}
\usepackage{comment}
\usepackage{natbib}
\usepackage{slashed}

\newcommand{\be}{\begin{equation}}
\newcommand{\e}{\end{equation}}
\newcommand{\beml}{\begin{subequations}}
\newcommand{\eml}{\end{subequations}}
\newcommand{\beq}{\begin{eqnarray}}
\newcommand{\eq}{\end{eqnarray}}
\newcommand{\ba}{\begin{array}}
\newcommand{\ea}{\end{array}}
\newcommand{\bpm}{\begin{pmatrix}}
\newcommand{\epm}{\end{pmatrix}}
\newcommand{\bc}{\begin{cases}}
\newcommand{\ec}{\end{cases}}

\usepackage{placeins}
\usepackage{float}
\usepackage{upgreek}

\usepackage{todonotes}

\usetikzlibrary{calc,intersections,through,hobby}
\usepackage{cleveref}

\usepackage{epsfig}
\newcommand{\AK}[1]{{\color{black}#1}}
\newcommand{\bigo}[1]{\oldcal{O}\left(#1\right)}
\DeclareMathAlphabet{\oldcal}{OMS}{cmsy}{m}{n}

\definecolor{amendments}{rgb}{0.0, 0.0, 0.7}

\usepackage[utf8]{inputenc}
\begin{document}
\title{Stability of \AK{$\varphi^4$-vector} model: four-loop $\varepsilon$ expansion study}

\author{L.\,Ts.\,Adzhemyan}
\affiliation{Saint Petersburg State University, 7/9 Universitetskaya Embankment, St. Petersburg, 199034 Russia}
\affiliation{Bogoliubov Laboratory of Theoretical Physics, Joint Institute for Nuclear Research, 6 Joliot-Curie, Dubna, Moscow region, 141980, Russian Federation}
\author{A.\,Kudlis}
\affiliation{ITMO University, Kronverkskiy prospekt 49, Saint Petersburg 197101, Russia}
\affiliation{Moscow Institute of Physics and Technology, Institutskiy Pereulok, 9, Dolgoprudnyi, Moscow Region 141701, Russia}

\begin{abstract}
\AK{In most papers, $\varphi^4$-field theory with the vector~($d$-component) field $\varphi_{\alpha}$ is considered as a particular case of the $n$-component field model for $n=d$ and $O(n)$ symmetry. However, in such a model the symmetry $O(d)$ admits an addition to the action of a term proportional to the squared divergence of the field $\sim h(\partial_{\alpha}\varphi_{\alpha})^2$. From the point of view of renormalization group analysis, it requires a separate consideration, because it may well change the nature of the critical behavior of the system.} Therefore, this frequently neglected term in the action requires a detailed and accurate study on the issue of existing of new fixed points and their stability. \AK{It is known that within the lower order of perturbation theory the only infrared stable fixed point with $h=0$} exists but the corresponding positive value of stability exponent $\omega_h$ is tiny. This led us to analyze this constant in higher orders of perturbation theory by calculating the 4-loop renormalization group contributions for $\omega_h$ in $d=4-2\varepsilon$ within Minimal Subtraction (MS) scheme, that should be enough to infer positivity or negativity of this exponent. The value turned out to be undoubtedly positive, although still small even in higher loops: $0.0156(3)$. These results cause that the corresponding term should be neglected in the action when analyzing the critical behaviour of $O(n)$-symmetric model. At the same time, the small value of $\omega_h$ shows that the corresponding corrections to the critical scaling are significant in a wide range.
\end{abstract}

\maketitle

\section{Introduction}
In recent years, the theory of critical behavior has experienced a renaissance. It is dictated by the emergence of new theoretical approaches~\cite{PhysRevD.86.025022,Cappelli2019,Simmons-Duffin2017,Kos2014,Kos2016} and by the fact that compared to couple decades ago a number of powerful numerical methods used for the needs of renormalization group (RG) approaches have appeared~\cite{BinothHeinrich:SectorDecomposition,Baikov2010mi,Lee2012mi,chetyrkin2017rstar,Batkovich2015rstar,Brown:TwoPoint,Panzer:HyperIntAlgorithms,BrownKreimer:AnglesScales,BatkovichKompanietsChetyrkin:6loop,KompanietsPanzer:LL2016,DUPUIS20211,PhysRevLett.123.240604,MOUHANNA2011}. Apart from that, the computational power of modern hardware opens up great prospects for breaking into high orders of perturbative critical thermodynamics by means of RG methods for that field models where previously it was unthinkable due to the technical features of calculations. One of these problems is the analysis of the field model, where the quadratic part of the action differs from the standard one by a term \AK{$\sim h(\partial_{\alpha}\varphi_{\alpha})^2$}. In the general case, such an additive, which preserves rotational symmetry, is not prohibited by any fundamental restrictions. To the surprise of the authors, a proper analysis of its presence in the action of the standard $O(n)$-symmetric model has not been presented in the literature, although this could potentially lead to a change the usual Heisenberg universality class. Moreover, the realization of such work is also motivated by the recently published paper of the authors, in which we analyze the action in the strong dipole-dipole coupling regime~\cite{KUDLIS2022115990}, where the term mentioned above was discarded. The neglecting of this additive was due to the belief that it does not change the universality class, giving as a stable fixed point one that contains the zero value of $h$ coordinate ($h^*=0$). \AK{Some discussion and the lowest-order analysis regarding the stability of such fixed point can be found in Ref.~\cite{PhysRevLett.30.559,PhysRevB.8.3349}. However, the smallness of the exponent requires checking for the fact that the sign of this quantity does not change in higher orders, as it frequently happens in the theory of critical behavior.}

The authors of this paper are sure that such omission in the general theory should be eliminated as soon as possible. Thus in this work, we analyze the stability of $h$-vector model within the MS scheme in the $4-2\varepsilon$ dimensions. On the basis of these results, we will give unambiguous conclusion regarding the stability of the corresponding fixed point, or rather, the stability of the ordinary Heisenberg point with respect to the presence of \AK{$\sim h(\partial_{\alpha}\varphi_{\alpha})^2$} term in the action.

The paper is organized as follows. In Sec.~\ref{sec:model_and_ren}, the model and renormalization scheme which we use in this work are described. Next, in Sec.~\ref{sec:num_res} the expansions for RG functions and numerical results are presented. At the end in Sec.~\ref{sec:conclusion}, we will draw a conclusion.

\section{Model and renormalization \label{sec:model_and_ren}}
The action we plan to study is as follows:
\begin{align} \label{eqn:full_act}
&S_0 = - \int d^d{x}\Bigg[\frac{1}{2} \left[(\partial_{\beta} \varphi_{0\alpha }) ^2+ m_0^2  \varphi_{0\alpha }^2 + h_0 (\partial_{\alpha}\varphi_{0\alpha})^2  \right]\nonumber\\
&\quad\qquad\qquad\qquad+ \frac{1}{4!} u_{0}T^{\alpha\beta\gamma\delta}_{\textup{S}}\varphi_{0\alpha} \varphi_{0\beta}  \varphi_{0\gamma} \varphi_{0\delta} \Bigg],
\end{align}
where $\varphi_{0\alpha}$ is $d$-component bare field, $u_{0}$ is bare coupling constants, $m_0^2$ is bare mass being proportional to $T-T_c$, where $T_c$ is the mean-field critical temperature. The tensor factor $T_{\textup{S}}^{\alpha\beta\gamma\delta}$ reads as follows:
\begin{align}
&T^{\alpha\beta\gamma\delta}_{\textup{S}} = \frac{1}{3}(\delta_{\alpha \beta} \delta_{\gamma \delta} +\delta_{\alpha \gamma} \delta_{\beta \delta}+\delta_{\alpha \delta} \delta_{\gamma \beta}).
\end{align}
The corresponding propagator within the momentum representation is
\begin{align}
G_{\alpha\beta}(\boldsymbol{p})=\dfrac{\delta_{\alpha\beta}-\hat{p}_{\alpha}\hat{p}_{\beta}}{p^2+m_0^2}+\dfrac{\hat{p}_{\alpha}\hat{p}_{\beta}}{(1+h_0)p^2+m_0^2},
\end{align}
where orientation vector defines as $\hat{p}_{\alpha}=p_{\alpha}/p$. \AK{The propagator denominator makes it easy to understand the region of stability, which is determined by the condition $h_0>-1$.} From the computational point of view, it is convenient to resort to projectors, in terms of which the propagator is written as:
\begin{align}\label{eqn:two_part_prop}
G_{\alpha\beta}(\boldsymbol{p})=\dfrac{P_{\alpha\beta}^{\bot}}{p^2+m_0^2}+\dfrac{P_{\alpha\beta}^{\parallel}}{(1+h_0)p^2+m_0^2},
\end{align}
with $P_{\alpha\beta}^{\bot}=\delta_{\alpha\beta}-\hat{p}_{\alpha}\hat{p}_{\beta}$ and $P_{\alpha\beta}^{\parallel}=\hat{p}_{\alpha}\hat{p}_{\beta}$.
Let us discuss the renormalization of the 
action~\eqref{eqn:full_act} within the $\varepsilon$ expansion technique. As is well known,
the only two- and four-point one-irreducible Green functions are spoiled by the pole singularities in $\varepsilon$. In the considered model, an additional divergence compared to the $O(n)$-symmetric model appears in the quadratic contribution $\sim h_0 p_{\alpha} p_{\beta}$. The counterterm necessary for the renormalization is reproduced by the multiplicative renormalization of the parameter $h_0$. This preserves the general multiplicative renormalizability of the theory. The renormalization procedure is reduced to introducing the necessary renormalization constants to bare field theory~\eqref{eqn:full_act}. The renormalized action has the following form:
\begin{align}\label{eqn:action_ren}
&S = - \int d^d{x}\Bigg[\frac{1}{2} \left[Z_1(\partial_{\beta} \varphi_{\alpha }) ^2+ Z_2 m^2  \varphi_{\alpha }^2 + Z_4 h (\partial_{\alpha}\varphi_{\alpha})^2  \right]\nonumber\\
&\quad\qquad\qquad\qquad+ \frac{1}{4!}Z_3\mu^{2\varepsilon} u T^{\alpha\beta\gamma\delta}_{\textup{S}}\varphi_{\alpha} \varphi_{\beta}  \varphi_{\gamma} \varphi_{\delta} \Bigg].
\end{align}
The renormalization constants $Z_i$ are calculated perturbatively in the form of series in renormalized coupling constant $u$ and for $u=0$ they satisfy $Z_i=1$. The transition from bare action~\eqref{eqn:full_act} to renormalized one~\eqref{eqn:action_ren} can be interpreted as redefining of field and system parameters from bare values to renormalized ones:
\begin{align}
     \varphi_{0\alpha}=\varphi_{\alpha}Z_{\varphi},\, m_0^2=m^2 Z_{m^2}, \, h_0=h Z_{h}, \, u_0=\mu^{2\varepsilon}u Z_{u}. \nonumber  
\end{align}
Comparing the bare~\eqref{eqn:full_act} and renormalized~\eqref{eqn:action_ren} action, the different renormalization constants are related to each other as:
\begin{align}\label{eqn:asd}
Z_1=Z_{\varphi}^2,\, Z_2=Z_{\varphi}^2 Z_{m^2}, \, Z_3=Z_u Z_{\varphi}^4, \, Z_4=Z_h Z_{\varphi}^2.
\end{align}
\AK{
Note that in Ref.~\cite{PhysRevB.43.833}, the difference between the renormalization constants $Z_1$ and $Z_4$ was interpreted as the difference between the renormalization constants of the transverse $\varphi_{\alpha}^{\perp}=P^{\perp}_{\alpha\beta}\varphi_{\beta}$ and longitudinal $\varphi_{\alpha}^{\parallel}=P^{\parallel}_{\alpha\beta}\varphi_{\beta}$  components of the field : $\varphi_{0\alpha}^{\parallel}=Z_{\varphi}^{\parallel}\varphi_{\alpha}^{\parallel}$ and $\varphi_{0\alpha}^{\perp}=Z_{\varphi}^{\perp}\varphi_{\alpha}^{\perp}$. However, in this case it is impossible to interpret the renormalization constant $Z_3$ in the form $Z_3=Z_u Z_{\varphi}^4$ that violates the multiplicative renormalizability of the model.}

In the MS scheme, the coefficients of the series in $u$ of the renormalization constants $Z_i$ do not depend on $m$ and $\mu$, but only on
dimensionless parameter $h$ which plays the role of a nonperturbative charge. The renormalization group equations for the Green functions can be obtained from the condition that their bare counterparts -- $G^{(0)}_n=\braket{\varphi_{0\alpha_1}\varphi_{0\alpha_2}\dots \varphi_{0\alpha_n}}$ -- do not depend on the parameter $\mu$. Taking into account the relation between bare and renormalized Green functions $G_{n}^{(0)}=Z_{\varphi}^nG_{n}^{\textup{R}}$, we act on both sides of the equality by the operation $\tilde{D}_{\mu}=\mu\partial_{\mu}|_{u_0,h_0,m_0}$:
\begin{align}\label{eqn:RG_eq_1_0}
    \tilde{D}_{\mu}G^{\textup{R}}_{n}+n\gamma_{\varphi} G_n^{\textup{R}}=0,
\end{align}
where $\gamma_{\varphi}=\tilde{D}_{\mu}\ln Z_{\varphi}$. Passing completely in the operator $\tilde{D}_{\mu}$ to the renormalized parameters:
\begin{align}
    \tilde{D}_{\mu}=D_{\mu}+(\tilde{D}_{\mu}m^2)\partial_{m^2}+(\tilde{D}_{\mu}h)\partial_{h}+(\tilde{D}_{\mu}u)\partial_{u},
\end{align}
we obtain the following equation for renormalized Green function:
\begin{align}\label{eqn:RG_eq_2_0}
    \Big[D_{\mu}+(\tilde{D}_{\mu}m^2)\partial_{m^2}+\beta_{h}\partial_{h}+\beta_{u}\partial_{u}+n\gamma_{\varphi}\Big]G^{\textup{R}}_{n}=0,
\end{align}
where the $\beta$-functions are defined as:
\begin{align}\label{eqn:beta_functions}
    \beta_u=\tilde{D}_{\mu}u, \quad \beta_h=\tilde{D}_{\mu}h.
\end{align}
Also, from the connection between bare and renormalized parameters, one can extract the following necessary relations:
\begin{gather}
\tilde{D}_{\mu}m^2=-\gamma_{m^2}m^2,\quad \beta_h=\tilde{D}_{\mu}h=-\gamma_{h}h,\nonumber\\ 
\beta_u=\tilde{D}_{\mu}u=-u(2\varepsilon+\gamma_u),\label{eqn:gammas_and_betas}
\end{gather}
where anomalous dimensions read as:
\begin{align}
\gamma_{m^2}=\tilde{D}_{\mu}\ln Z_{m^2},\quad \gamma_{h}=\tilde{D}_{\mu}\ln Z_{h}, \gamma_{u}=\tilde{D}_{\mu}\ln Z_{u}.
\end{align}
Taking into account all the above, we get the following RG equation:
\begin{equation}\label{eqn:RG_eq_3_0}
    \Big[\mu\partial_{\mu}-\gamma_{m^2}m^2\partial_{m^2}+\beta_{h}\partial_{h}+\beta_{u}\partial_{u}+n\gamma_{\varphi}\Big]G^{\textup{R}}_{n}=0.
\end{equation}
The anomalous dimensions $\gamma_{\varphi}$, $\gamma_{m^2}$, $\gamma_{h}$, and $\gamma_{u}$ do not possess poles in $\varepsilon$, moreover within MS scheme they depend only on $u$ and $h$ and do not depend on $\varepsilon$ at all:
\begin{equation}\label{eqn:RG_eq_4_0}
    \gamma_{i}=(\beta_{h}\partial_{h}+\beta_{u}\partial_{u})\ln Z_i.
\end{equation}
Thus, all RG-functions are expressed in terms of renormalization constants, except the trivial dependence of $\beta_u$ on $\varepsilon$. Taking into account the relations~\eqref{eqn:RG_eq_4_0} and~\eqref{eqn:gammas_and_betas} the beta functions can be found from the following system:
\begin{align}\label{eqn:RG_eq_5_0}
   &\beta_u=-u\left[2\varepsilon+ \left(\beta_u\partial_{u}+\beta_h\partial_{h}\right)Z_u\right],\\
   &\beta_h=-h\left[\left(\beta_u\partial_{u}+\beta_h\partial_{h}\right)\AK{Z_h}\right],
\end{align}
The critical behavior of the system is determined by the fixed point, or, equivalently, by the zeros of the $\beta$-functions:
\begin{align}\label{eqn:beta_functions_zeros}
\beta_u(u^*,h^*)=0, \quad \beta_h(u^*,h^*)=0.
\end{align}
If the analyzed fixed point is infrared-stable that is determined by the behavior of the $\beta$-functions then in the critical region the Green function satisfies the following equation:
\begin{equation}\label{eqn:RG_eq_6_0}
    \Big[\mu\partial_{\mu}-\gamma_{m^2}^*m^2\partial_{m^2}+n\gamma_{\varphi}^*\Big]G^{\textup{R}}_{n}=0, \, \gamma_i^*=\gamma_i(u^*,h^*).
\end{equation}
In the next section, we will apply these formulas to specific Feynman diagrams.

\section{RG expansions and numerical estimates}\label{sec:num_res}
Since we are analyzing the stability of an $O(n)$-symmetric fixed point, which, as we will show, is stable with respect to a perturbation of the type \AK{$\sim h(\partial_{\alpha}\varphi_{\alpha})^2$}, we do not need to renormalize the mass to obtain expressions for the critical exponents $\nu$, $\alpha$, etc, they will be the same as in case of $O(n)$-symmetric universality class with $n=4-2\varepsilon$. Moreover, to simplify the calculations, one can resort to a massless computational scheme. Let us write down the way we calculate the renormalization constants. As was said, due to multiplicative renormalizability of the model it is enough to remove divergences in two- and four-point one-particle irreducible Green functions which can be denotes as $\Gamma^{(2)}_{\alpha\beta}$ and $\Gamma^{(4)}_{\alpha\beta\gamma\delta}$ respectively. The counterterms eliminating the divergences in these functions are polynomials in external momenta. \AK{For $\Gamma_{\alpha\beta}^{(2)}$ it is quadratic polynomial in the form $c_1(\varepsilon) p^2 \delta_{\alpha\beta} + c_2(\varepsilon) p_{\alpha}p_{\beta}=p^2(c_1(\varepsilon)\delta_{\alpha\beta} + c_2(\varepsilon) \hat{p}_{\alpha}\hat{p}_{\beta})$, while for $\Gamma^{(4)}_{\alpha\beta\gamma\delta}$ the polynomial is of the zero order. In addition, from the point of view of convenience, we will use the $\overline{\text{MS}}$ scheme which is expressed by the transition to another perturbative charge $v$:
\begin{align}
 u = v (4\pi)^{d/2} e^{(2-d/2)\gamma},   
\end{align}
where $\gamma$ is the Euler constant. Thus, we will determine the renormalization constants based on the finiteness condition for the functions $\partial_{p^2}\Gamma_{\alpha\beta}^{(2)}$ and $\Gamma^{(4)}_{\alpha\beta\gamma\delta}/v\mu^{2\varepsilon}$:
\begin{align}\label{eqn:cond_renorm_1}
&\left.\frac{\partial\Gamma_{\alpha\beta}^{(2)}}{\partial p^2}\right.=\delta_{\alpha\beta}Z_1\left[1-\left(\frac{Z_3}{Z_1^2}\right)^2 v^2 \left(\frac{\mu}{p}\right)^{4\varepsilon}\vcenter{\hbox{
                                    \begin{tikzpicture}[use Hobby shortcut, scale=0.8]
                                      \draw (-0.45,0) .. (0.0,0.35) .. (0.45,0);
                                      \draw (-0.45,0) .. (0.0,-0.35) .. (0.45,0);
                                      \draw (-0.55,0) -- (0.55,0);
                                      \fill (-0.45,0) circle (1pt);
                                      \fill (0.45,0) circle (1pt);
                                    \end{tikzpicture}
                                    }}_{\delta}+\dots\right]\nonumber\\
&+\hat{p}_{\alpha}\hat{p}_{\beta}h Z_4\left[1-\left(\frac{Z_1}{Z_4 h }\right)\left(\frac{Z_3}{Z_1^2}\right)^2 \left(\frac{\mu}{p}\right)^{4\varepsilon}v^2\vcenter{\hbox{
                                    \begin{tikzpicture}[use Hobby shortcut, scale=0.8]
                                      \draw (-0.45,0) .. (0.0,0.35) .. (0.45,0);
                                      \draw (-0.45,0) .. (0.0,-0.35) .. (0.45,0);
                                      \draw (-0.55,0) -- (0.55,0);
                                      \fill (-0.45,0) circle (1pt);
                                      \fill (0.45,0) circle (1pt);
                                    \end{tikzpicture}
                                    }}_{\hat{p}}+\dots\right],\\
&\left.-\frac{\Gamma^{(4)}_{\alpha\beta\gamma\delta}}{v\mu^{2\varepsilon}}\right.=T_{\textup{S}}^{_{\alpha\beta\gamma\delta}}Z_3\left[1-\left(\frac{Z_3}{Z_1^2}\right)v \left(\frac{\mu}{p}\right)^{2\varepsilon}\vcenter{\hbox{
                                     \begin{tikzpicture}[use Hobby shortcut, scale=0.9]
                                       \draw (-0.35,-0.1) .. (90:0.15) .. (0.35,-0.1);
                                       \draw (-0.35,0.1) .. (-90:0.15) .. (0.35,0.1);
                                       \fill (-0.3,0) circle (1pt);
                                       \fill (0.3,0) circle (1pt);
                                     \end{tikzpicture}
                                     }} +\dots  \right],\label{eqn:cond_renorm_2}
\end{align}
where symbols $\delta$ and $\hat{p}$ denote the contributions of corresponding tensor structures, the multiplier $p^{-n\varepsilon}$ appears when the change of integration variable from momentum to dimensionless (in units of external momentum) one is performed. In this case the left diagrams depend only on $\varepsilon$ and $h$. Having done all of these steps, the obtained expressions for renormalization constants have the following structure:
\begin{align}
    Z_i=1+\sum\limits_{j=1}u^j\sum\limits_{l=1}^jc_{jl}^{(i)}(h)\varepsilon^{-l}.
\end{align}}
\begin{table}[t!]
\centering
\caption{Two-point Feynman diagrams up to four loops which should be calculated in order to extract the $\varepsilon$ expansion for $\omega_h$. Each graph is accompanied by Nickel-index. Nickel-index is commonly used in describing the topologies of Feynman
diagrams. The explanation of its modern modifications was described in
detail in Ref.~\cite{bkkn2014}.}
\label{tab:two_leg_diag}
\setlength{\tabcolsep}{1.4pt}
\renewcommand{\arraystretch}{1.5}
\begin{tabular}{c}
$\ \   \ e111|e| \qquad\quad \  e112|22|e| \qquad\qquad\quad  \   e112|33|e33|| \qquad \ \ $\\
$\vcenter{\hbox{\begin{tikzpicture}[use Hobby shortcut, scale=0.8]
                                      \draw (-0.8,0) .. (0.0,0.6) .. (0.8,0);
                                      \draw (-0.8,0) .. (0.0,-0.35) .. (0.8,0);
                                      \draw (-0.95,0) -- (0.95,0);
                                      \fill (-0.8,0) circle (1pt);
                                      \fill (0.8,0) circle (1pt);
                                      \draw (1.1,0) -- (1.25,0);
                                      \draw (1.25,0) .. (2.05,-0.35) .. (2.85,0);
                                      \draw (1.25,0) .. (2.05,0.25) .. (2.85,0);
                                      \draw (2.85,0) .. (3.65,0.25) .. (4.45,0);
                                      \draw (2.85,0) .. (3.65,-0.35) .. (4.45,0);
                                      \draw (1.25,0) .. (2.85,0.6) .. (4.45,0);
                                      \draw (4.45,0) -- (4.6,0);
                                      \fill (1.25,0) circle (1pt);
                                      \fill (2.85,0) circle (1pt);   
                                      \fill (4.45,0) circle (1pt);
                                      \draw (4.75,0) -- (4.9,0);
                                      \draw (4.9,0) .. (5.6,-0.35) .. (6.3,0);
                                      \draw (4.9,0) .. (5.6,0.22) .. (6.3,0);
                                      \draw (6.3,0) .. (7,0.22) .. (7.7,0);
                                      \draw (6.3,0) .. (7,-0.35) .. (7.7,0);
                                      \draw (7.7,0) .. (8.4,-0.35) .. (9.1,0);
                                      \draw (7.7,0) .. (8.4,0.22) .. (9.1,0);
                                      \draw (4.9,0) .. (7,0.6) .. (9.1,0);
                                      \draw (4.45,0) -- (4.6,0);
                                      \draw (9.1,0) -- (9.25,0);
                                      \fill (4.9,0) circle (1pt);
                                      \fill (6.3,0) circle (1pt);   
                                      \fill (7.7,0) circle (1pt);
                                      \fill (9.1,0) circle (1pt);
\end{tikzpicture}}}$\\
$  \qquad e112|23|33|e| \qquad  e112|e3|333||\qquad \quad e123|e23|33|| \qquad \quad $\\
$\vcenter{\hbox{\begin{tikzpicture}[use Hobby shortcut, scale=0.8]
                                      \draw (0.0,0.8) .. (0.32,0.0) .. (0.0,-0.8);
                                      \draw (0.0,0.8) .. (-0.32,0.0) .. (0.0,-0.8);
                                      \draw (1.6,0.8) .. (1.92,0.0) .. (1.6,-0.8);
                                      \draw (1.6,0.8) .. (1.27,0.0) .. (1.6,-0.8);
                                      \draw (0.0,0.8) -- (1.6,-0.8);
                                      \draw (0.0,0.8) -- (1.6,0.8);
                                      \draw (0.0,-0.8) -- (1.6,-0.8);
                                      \draw (0.0,-0.8) -- (-0.25,-0.8);
                                      \draw (1.6,0.8) -- (1.85,0.8);      
                                        \fill (0.0,0.8) circle (1pt);
                                        \fill (0.0,-0.8) circle (1pt);
                                        \fill (1.6,0.8) circle (1pt);
                                        \fill (1.6,-0.8) circle (1pt);
                                      \draw (2.82,0.8) .. (2.5,0.0) .. (2.82,-0.8);
                                      \draw (2.82,0.8) .. (3.14,0.0) .. (2.82,-0.8);
                                      \draw (4.82,0.8) .. (4.5,0.0) .. (4.82,-0.8);
                                      \draw (4.82,0.8) .. (5.14,0.0) .. (4.82,-0.8);
                                      \draw (2.82,0.8) -- (4.82,0.8);
                                      \draw (2.82,-0.8) -- (4.82,-0.8);   \draw (4.84,0.8) -- (4.84,-0.8);
                                      \draw (2.82,-0.8) -- (2.55,-0.8);
                                      \draw (2.82,0.8) -- (2.55,0.8);  

                                      \fill (4.84,0.8) circle (1pt);
                                      \fill (4.84,-0.8) circle (1pt);
                                      \fill (2.82,0.8) circle (1pt);
                                      \fill (2.82,-0.8) circle (1pt);     \draw (5.8,-0.8) -- (8.8,-0.8);
                                    \draw (5.8,-0.8) ..(6.25,0.0) ..  (7.3,0.8);
                                    \draw (8.8,-0.8) ..(8.35,0.0) ..  (7.3,0.8);
                                    \draw (5.8,-0.8) -- (7.3,-0.3);
                                    \draw (8.8,-0.8) -- (7.3,-0.3);
                                    \draw (7.3,-0.3) .. (7.6,0.25) .. (7.3,0.8);
                                    \draw (7.3,-0.3) .. (7.,0.25) .. (7.3,0.8);
                                    \draw (8.8,-0.8) -- (8.95,-0.8);
                                    \draw (5.8,-0.8) -- (5.65,-0.8);
                                     \fill (8.8,-0.8) circle (1pt);
                                     \fill (5.8,-0.8) circle (1pt); 
                                     \fill (7.3,-0.3) circle (1pt);
                                     \fill (7.3,0.8) circle (1pt);
\end{tikzpicture}}}$
\end{tabular}
\end{table}
\AK{It is essential that the pole contributions, which are obtained in~\eqref{eqn:cond_renorm_1} and~\eqref{eqn:cond_renorm_2} from the product of the expansion $\left(\mu/p\right)^{k\varepsilon}=1+k \varepsilon \ln{\mu/p}\dots$ and higher-order pole contributions (from diagrams and renormalization constants) cancel each other, which is a very important consequence of the renormalizability of the theory. This leads to the fact that the renormalization constants do not depend on $\mu/p$. At the same time, the interesting for us contribution in~\eqref{eqn:cond_renorm_2} does not depend on the way of momentum passing through a diagram.}

Having obtained $Z_1$, $Z_3$, and $Z_4$, as functions of $v$, $h$, and $\varepsilon$ we can extract $Z_{\varphi}$, $Z_{v}$, and $Z_{h}$ by means of the following expressions:
\begin{align}
Z_v = \dfrac{Z_3}{Z_1^2}, \ Z_h = \dfrac{Z_4}{Z_1}, \ Z_{\varphi} = \sqrt{Z_1}.
\end{align}


Let us compose all the expressions, as a first step we need to find the lowest order for the $\beta$-function of the nonperturbative charge $h$ in order to analyze a possibility to obtain a new fixed point or to make sure that only zero-valued $h$ is possible. For this purpose, it is enough to calculate the pole of the diagram $e111|e|$\footnote{Originally description of this diagram notation was presented in the following report: 
\url{http://users.physik.fu-berlin.de/~kleinert/nickel/guelph.pdf}}:
\begin{align}
\vcenter{\hbox{\begin{tikzpicture}[use Hobby shortcut, scale=0.8]
                                      \draw (-0.45,0) .. (0.0,0.35) .. (0.45,0);
                                      \draw (-0.45,0) .. (0.0,-0.35) .. (0.45,0);
                                      \draw (-0.55,0) -- (0.55,0);
                                      \fill (-0.45,0) circle (1pt);
                                      \fill (0.45,0) circle (1pt);
                                    \end{tikzpicture}
                                    }}_{\alpha\beta}&=-\dfrac{h_0 (20 + 36 h_0 + 19 h_0^2)}{432(1 + h_0)^3 \varepsilon}
   \hat{p}_{\alpha}\hat{p}_{\beta}\\
&\quad\quad-\dfrac{(36 + 76 h_0 + 63 h_0^2 + 20 h_0^3)}{432 (1 + h_0)^3\varepsilon}\delta_{\alpha\beta}+\bigo{\varepsilon^0}\nonumber\\
&=\vcenter{\hbox{
                                    \begin{tikzpicture}[use Hobby shortcut, scale=0.8]
                                      \draw (-0.45,0) .. (0.0,0.35) .. (0.45,0);
                                      \draw (-0.45,0) .. (0.0,-0.35) .. (0.45,0);
                                      \draw (-0.55,0) -- (0.55,0);
                                      \fill (-0.45,0) circle (1pt);
                                      \fill (0.45,0) circle (1pt);
                                    \end{tikzpicture}
                                    }}_{\hat{p}}\hat{p}_{\alpha}\hat{p}_{\beta}+\vcenter{\hbox{
                                    \begin{tikzpicture}[use Hobby shortcut, scale=0.8]
                                      \draw (-0.45,0) .. (0.0,0.35) .. (0.45,0);
                                      \draw (-0.45,0) .. (0.0,-0.35) .. (0.45,0);
                                      \draw (-0.55,0) -- (0.55,0);
                                      \fill (-0.45,0) circle (1pt);
                                      \fill (0.45,0) circle (1pt);
                                    \end{tikzpicture}
                                    }}_{\delta}\delta_{\alpha\beta}+\bigo{\varepsilon^0},\nonumber
\end{align}
where we have taken into account symmetry and combinatorial factors~($n$ everywhere is replaced by $d=4-2\varepsilon$). After substituting $h_0=hZ_4/Z_1$, by means of Eqns.~\eqref{eqn:cond_renorm_1},~\eqref{eqn:cond_renorm_2}  and simplified (valid only in low order) expression for $\beta$-function of $h$ ($\beta_h=2\varepsilon h v\partial_{v}\ln{Z_h}$), this diagram allows to obtain the following result:
\begin{align}
\beta_h=\frac{h \left(5 h^2+6 h+4\right)}{27 (h+1)^2}v^2+O(\AK{v^3}).
\end{align}
It is easy to see that for a fixed point equation here is only one real root -- $h=0$. At this stage, the important question connected with the stability of this point appears. If it is not stable, then it is necessary to look for other fixed points and, obviously, this can be done only in higher orders of perturbation theory. \AK{If the point is stable ($\omega_h>0$), but numerical value of the exponent is small, it is necessary to check whether the sign does not change in higher orders.} The latter is realized in our case -- as it turned out, $\omega_h>0$ does not become negative in higher orders. It is worth noting that there are situations in the literature when the conclusion regarding one or another class of universality changes with the growth of the orders of the perturbation theory~\cite{ADZHEMYAN2019332,KOMPANIETS2020114874}.

Taking into account the value of fixed point coordinate $v^*=\varepsilon/2$ which is the same as in case of $O(n)$-symmetric theory with $ n=4-2\varepsilon$, the corrections to scaling exponent equals to $\varepsilon^2/27$ which in case of physical value of $\varepsilon=1/2$ gives $\sim 0.009$. As was said above, such a small value of the correction to scaling exponent forces us to make sure that in higher orders of perturbation theory, the sign of this quantity does not change to negative. Note that since we are interested in the case $h = 0$, it will suffice to consider only terms linear in $h$ in the $\beta_h$ function. This fact allows one to calculate only linear contributions in $h$ in all two-legs diagrams, and four-legs can be computed initially for $h=0$. In the future, only this consideration will be enough, since the fixed point with zero $h$ value turns out to be stable. The corresponding topologies of diagrams are presented in Table~\ref{tab:two_leg_diag}.

Thus, having calculated all necessary up to four-loops diagrams, we obtain the following expansions for $\beta$-functions :
\begin{align}
\beta_v&=-2\varepsilon v +4 v^2 -\dfrac{26}{3}v^3+ \dfrac{\left(6984 + 4032\zeta_3 \right)}{216}v^4 + O(v^5),\nonumber\\
\beta_h&=\dfrac{4h}{27}v^2 -\dfrac{4h}{27}v^3+\dfrac{2344 h}{2187}v^4 + O(v^5).\label{eqn:beta_h_model}
\end{align}
It is interesting to note that $\beta_v$ differs from  its $O(4)$-symmetric counterpart~\cite{KP17} starting only from $v^4$:
\begin{align}
\beta_v^{4}=-2\varepsilon v +4 v^2 -\dfrac{26}{3}v^3+ \dfrac{\left(7176 + 4032\zeta_3 \right)}{216}v^4 + O(v^5),\nonumber
\end{align}
The fixed point extracted from~\eqref{eqn:beta_h_model} has the following coordinates: 
\begin{align}\label{eqn:fixed_coord}
&v^*=\dfrac{1}{2}\varepsilon+\dfrac{13}{24}\varepsilon^2+\dfrac{(47-168\zeta_3)}{288}\varepsilon^3+O(\varepsilon^4),\nonumber\\
&h^*=0.
\end{align}
Let us now calculate the $\varepsilon$ expansion for the correction to scaling exponent $\omega_h=\partial_{h}\beta_h(h^*,v^*)$:
\begin{align}\label{eqn:eps_expansion_omega_h}
    \omega_h=\dfrac{1}{27}\varepsilon^2+\dfrac{5}{81}\varepsilon^3+\dfrac{(2605-3024 \zeta_3)}{34992}\varepsilon^4+O(\varepsilon^5),
\end{align}
which numerically reads as follows:
\begin{align}\label{eqn:num_exapnsion_omegah}
    \omega_h=0.03704\varepsilon^2+0.06173 \varepsilon^3-0.02944 \varepsilon^4+O(\varepsilon^5).
\end{align}

In order to obtain the proper numerical estimates for $\omega_h$ the different resummation strategies should be applied~\cite{PhysRevD.15.1544}. Before we get into the resummation procedures, let us look at the estimate that can be obtained by a simple direct summation in case of $\varepsilon=1/2$. This gives $\sim 0.0151$. Later we will see that due to the favorable structure of the expansion, such an estimate turns out to be quite close to the numbers that are obtained using various resummation techniques. 

As the first method, we choose the simple Pad\'e approximants. Let the analyzed expansion for $\omega_h$ reads as $\varepsilon^2\sum_{k=0}^{2}c_k\varepsilon^k$. The standard step that should be done in this case is to reduce the polynomial to such a form which starts with the constant. The corresponding highest available approximant is $P_{[1/1]}$:
\begin{align}
\textup{P}_{[1/1]}\left[\sum_{k=0}^{2}c_k\varepsilon^k\right]=\dfrac{0.037037+0.0793899 \varepsilon}{1+0.476861 \varepsilon}.
\end{align}
Keeping in mind the factor $\varepsilon^2$, for physically interesting case ($\varepsilon=1/2$) we have $0.01549$. The number of digits we have left is due to the accuracy of alternative resummation methods.

\begin{figure}[t!]
    \centering
    \includegraphics[width=1\linewidth]{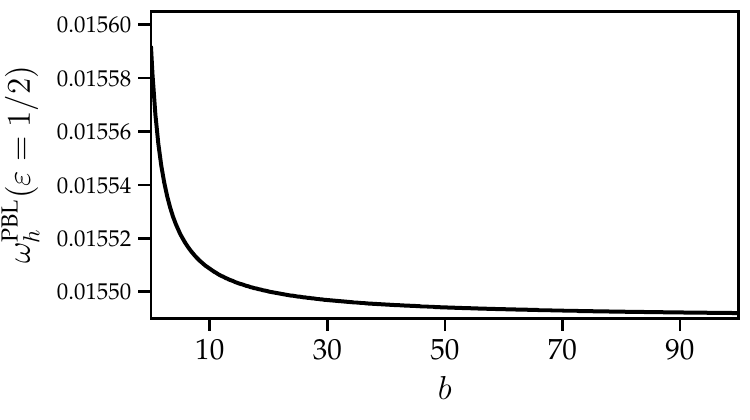}
    \caption{Dependence of constructed by means of PBL technique numerical estimate of exponent $\omega_h$ on the value of fitting parameter $b$. The corresponding Pad\'e approximant is $[1/1]$.}
    \label{fig:pbl_est}
\end{figure}

The more tricky method is the Pad\'e-Borel-Leroy (PBL) technique. This approach is based on the so called Borel-transformation. The resummed value can be found by means of the following formula:
\begin{align}
\omega_{h,b}^{\textup{PBL}}(\varepsilon) =\varepsilon^2\int\limits_{0}^{\infty}dt^{} \, t^{b} e^{-t}P_{[1/1]}\left[\sum\limits_{k=0}^{2}\dfrac{c_k \left(\varepsilon t\right)^k}{\Gamma(k+1+b)} \right].
\end{align}
The fitting parameter $b$ is chosen from the principle of maximum or minimum value of dependence of analyzed quantity on $b$ variation. The dependence of numerical estimate of PBL approximant on parameter $b$ is demonstrated in Fig.~\ref{fig:pbl_est}. As an error, we will take the variation of the change in the PBL estimate when changing the fitting parameter over the entire range $[0,\infty)$. Thus, our PBL estimate for $\omega_h$ equals to $0.01559(10)$.

A more advanced approach, called the conformal Borel resummation technique, still uses the Borel transform, but the analytic continuation is sought using a special conformal mapping. The general idea can be formulated as follows. Let the asymptotic behavior of the series coefficients of analyzed expansion $c(\varepsilon)=\sum c_k\varepsilon^k$ for large order behave as
\begin{align}
\label{eqn:hoab_A_expansion}
c_k \xrightarrow[k \rightarrow \infty]{} const\, k! k^{b^{}_0}(-a)^k ,
\end{align}
where $1/a$ is the radius of convergence and $b^{}_0$ is fixed by the high-order asymptotic behavior of the series. The authors of Ref.~\cite{KP17} proposed to treat the parameter $b=b^{}_0+3/2$ as a free one, to be determined variationally. Once a Borel transformation, based on this modified asymptotic form, is performed, the variable $\varepsilon$ is conformally mapped onto $w$ as:
\begin{align}
w(\varepsilon)=\dfrac{\sqrt{1+a\varepsilon}-1}{\sqrt{1+a\varepsilon}+1}, \quad \varepsilon(w)=\dfrac{4w}{a(1-w)^2}.
\end{align}
Moreover, it is assumed that the expansion has the strong asymptotic behavior $c(\varepsilon)\sim \epsilon^\lambda,\ \ \varepsilon\rightarrow \infty$. The results are then improved by a preliminary homogeneous homographic transformation,
\begin{align}
\label{eqn:homographic_trans}
\varepsilon(\varepsilon')  \rightarrow \dfrac{\varepsilon'}{1+q\varepsilon'}, \quad \varepsilon'(\varepsilon)  \rightarrow \dfrac{\varepsilon}{1-q\varepsilon},
\end{align}
and the final approximate estimates are found by applying the steps mentioned above to the new $\varepsilon'$ expansion.
At the end, the optimal parameters are chosen such that the final estimate is the least sensitive to changes in these parameters in the vicinity of their optimal values. After analyzing the numerical values of the estimates obtained using this method, we managed to come to the following number: $0.0156(12)$.
\begin{table}[t]
\centering
\caption{Numerical estimates of the correction to scaling exponents $\omega_h$ obtained by means of different resummation strategies. An error is indicated in those cases where it is implied by the resummation algorithm.} %
\label{tab:omega_h}
\setlength{\tabcolsep}{6.1pt}
\begin{tabular}{cccc}
\hline
\hline
Pad\'e & Pad\'e-Borel-Leroy & Conformal Borel & SSFT \\
\hline
$0.01549$ &    $0.01559_{b=0}(10)$      &$0.0156(12)$      &$0.01577$   \\
\hline
\hline
\end{tabular}
\end{table}

The last procedure we use is the one based on self-similar factor transformations (SSFT)~\cite{Yukalov2022}. The essence of the technique is to replace the summation with the product, that should improve the convergence.
\begin{align}
\omega_{\textup{SSFT}}(\varepsilon)=c_0\varepsilon^2 \prod\limits_{j=1}^{1}(1+A_j)^{n_j}, 
\end{align}
where constants $A_j$ and $n_j$ can be found from:
\begin{align}
\sum_{k=0}^{2}(c_k/c_0)\varepsilon^k= \prod\limits_{j=1}^{1}(1+A_j)^{n_j}+\bigo{\varepsilon^3}.
\end{align}
Based on the expansion~\eqref{eqn:num_exapnsion_omegah} we obtain $A_1=2.62039$ and $n_1=0.63604$. For $\varepsilon=1/2$ the value of exponent equals to $0.01577$. 

Numerical results that were obtained using various resummation techniques were collected in Table~\ref{tab:omega_h}. It can be seen that the fourth order of perturbation theory makes it possible to give a good grouping of answers. Based on the results obtained by means of different resummation strategies, we come to the following final estimate for correction to scaling exponent:
\begin{align}
\omega_h=0.0156(3).   
\end{align}
The resulting number allows us to make an affirmative conclusion regarding the stability of the standard Heisenberg  universality class to the appearance of the term $\sim h(\partial_{\alpha}\varphi_{\alpha})^2$.

\section{Conclusion}\label{sec:conclusion}
To sum up, in this paper we solved the problem, which for a long time remained without required attention -- the stability of an $O(n)$-symmetric fixed point with respect to the additional term proportional to the divergence of the field in the action. For this purpose, we have calculated four-loop $\varepsilon$ expansion for correction to scaling exponent $\omega_h$. In the lower orders, we have found that there is only one fixed point with zero $h$. Analyzing higher orders, we made sure that the correction to scaling exponent, although small, but is unambiguously positive. Thus, we can conclude that discarding the term $\sim h(\partial_{\alpha}\varphi_{\alpha})^2$ when considering the critical behavior of systems described by $O(n)$-symmetric theory is fully justified.

\begin{acknowledgments}
We gratefully acknowledge A. Pikelner for the help with diagrams. The work of A.K. was supported by Grant of the Russian
Science Foundation No 21-72-00108.
\end{acknowledgments}
\allowdisplaybreaks
\begin{widetext}
\appendix
\section{Alternative implementation of the renormalization procedure: fixed value of $n$ \label{app:alt_ren}}
If in all diagrams the combinatorial factors are considered for arbitrary $n$, then we obtain the following $\beta$-functions:
\begin{align}
    \beta_v&=-2 \varepsilon v +\dfrac{(n+8)}{3}v^2 - \dfrac{\left(3n+14\right)}{3}v^3 +\dfrac{\left(2112\zeta_3+2960+(480\zeta_3+922)n+33 n^2\right)}{216}v^4+\bigo{v^5} ,\\
\beta_h/h&=\Bigg[\dfrac{\left(n^2-8\right)}{18 (n-1)}\Bigg]v^2-\Bigg[\dfrac{24(n^2-16)}{648 \varepsilon (n-1)}-\dfrac{(4032+24 n-260 n^2-n^3)}{648 (n-1)}\Bigg]v^3+\Bigg[-\dfrac{n (n^2-16)}{324 \varepsilon^2(n-1)}\nonumber\\
&+\dfrac{(36480 - 34512 n - 4068 n^2 + 2140 n^3 + 107 n^4)}{3888 \varepsilon(n-1)^2} +\dfrac{1}{23328 (n-1)^2}(1940544+n(384\pi^2-1828224)\nonumber\\
&-n^2 (223384+384\pi^2)+n^3(115098-24\pi^2)+n^4(6303+24\pi^2)-47n^5)\Bigg]v^4 +\bigo{v^5},\label{eqn:second_beta_n_dep}
\end{align}
The corresponding fixed point has the following coordinates: 
\begin{align}
&v^*=\dfrac{6}{n+8}\varepsilon+\dfrac{36(3 n+14)}{(n+8)^3}\varepsilon^2+\dfrac{3\left(4544- 
 16896\zeta_3 + n (1760 - 5952\zeta_3) + n^2 (110 - 480 \zeta_3)-33 n^3\right)}{(n+8)^5}\varepsilon^3+\bigo{\varepsilon^4},\nonumber\\
&h^*=0.
\end{align}
From this step, we put $n=4-2\varepsilon$ and reexpand all series. The expansion for $v^*$ now reads as:
\begin{align}
&v^*=\dfrac{1}{2}\varepsilon+\dfrac{5}{8}\varepsilon^2+\dfrac{(85-168\zeta_3)}{288}\varepsilon^3+O(\varepsilon^4),
\end{align}
which differs from~\eqref{eqn:fixed_coord}, but fixed point coordinate itself has no physical sense. Having obtained the coordinate of fixed point, we can take $h$-derivative from~\eqref{eqn:second_beta_n_dep} and reexpand it substituting $n=4-2\varepsilon$:
\begin{align}
    \omega_h=\dfrac{1}{27}\varepsilon^2+\dfrac{5}{81}\varepsilon^3+\dfrac{(2605-3024 \zeta_3)}{34992}\varepsilon^4+O(\varepsilon^5),
\end{align}
which coincides with~\eqref{eqn:eps_expansion_omega_h}. We note an interesting fact that the pseudo-poles that appeared in the $\beta$-function~\eqref{eqn:second_beta_n_dep} are connected only with the fact that we tried to move away from the requirement that the dimensions of the field and space should coincide. Ultimately, these poles do not affect the final answer.
\end{widetext}
\bibliography{hmodel}
\end{document}